\documentstyle[aps,prl,multicol]{revtex} 
\input{epsf}
\begin{document}
\draft \title{Modulated Amplitude Waves and the \\
Transition from Phase to Defect Chaos 
}
\author{Lutz Brusch$^1$, Mart{\'\i}n G.
  Zimmermann$^{2,*}$, Martin van Hecke$^{1,3}$, Markus B\"ar$^1$, 
and Alessandro Torcini$^4$} 
\address{$^1$ Max-Planck-Institut f\"ur Physik komplexer
  Systeme, N\"othnitzer Stra{\ss}e 38, D-01187 Dresden, Germany}
  \address{$^2$ Instituto Mediterraneo de Estudios Avanzados, IMEDEA
  (CSIC-UIB), E-07071 Palma de Mallorca, Spain} \address{$^3$Center
  for Chaos and Turbulence Studies, The Niels Bohr Institute,
  Blegdamsvej 17, 2100 Copenhagen, Denmark} \address{$^4$
  Istituto Nazionale di Fisica
  della Materia, Unit\`a di Firenze, Largo Enrico Fermi 2, I-50125
  Firenze, Italy} 
\date{\today} 
\maketitle

\begin{abstract}
The mechanism for transitions from phase to defect chaos in the
one-dimensional complex Ginzburg-Landau equation (CGLE) is presented. We
introduce and describe periodic coherent structures of the CGLE, called
Modulated Amplitude Waves (MAWs). MAWs of various period $P$ occur naturally
in phase chaotic states. A bifurcation study of the MAWs reveals that for
sufficiently large period, pairs of MAWs cease to exist via a saddle-node
bifurcation. For periods beyond this bifurcation, incoherent near-MAW
structures occur which evolve toward defects.  This leads to our main result:
the transition from phase to defect chaos takes place when the periods of MAWs
in phase chaos are driven beyond their saddle-node bifurcation.
\end{abstract}

\begin{figure} \vspace{-0.1cm}
 \epsfxsize=1.\hsize \mbox{\hspace*{-.06 \hsize} \epsffile{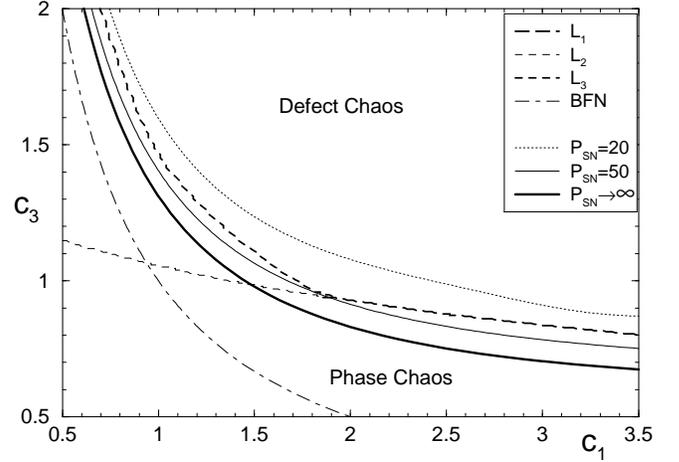} }
\caption[]{Phase diagram of the CGLE showing the BFN, $L_1$, $L_2$ and $L_3$
transitions (after \cite{chate}). Between the $L_2$ and $L_3$ curves, there is
the hysteretic regime where either phase or defect chaos can occur; in the
latter case, defects persist up to the $L_2$ transition. Notice how the $L_1$
and $L_3$ transitions to defect chaos lie above our lower ($P\rightarrow
\infty$) bounds. Also shown are the SN locations for $P\!=\!20,50$.
}\label{figPD}
\end{figure}

\begin{multicols}{2}  
\narrowtext
Spatially extended systems can exhibit, when driven away from equilibrium,
irregular behavior in space and time: this phenomenon is commonly referred to
as {\em spatio-temporal} chaos \cite{CH}.  The one-dimensional 
complex Ginzburg-Landau equation (CGLE):
\begin{equation} 
  \partial_{t} A = A + (1+ ic_1) \partial_{x}^{2} A 
- (1-i c_3) |A|^2 A ~ , \label{cgle}
\end{equation} 
describes pattern formation near a Hopf bifurcation and has become a popular
model to study spatiotemporal chaos
\cite{CH,janiaud,SupEck,KS,chao1,bazhenov,chate,saka,egolf,miguel,torcini,saar,mvh}.
As a function of $c_1$ and $c_3$, the CGLE exhibits two qualitatively
different spatiotemporal chaotic states known as phase chaos (when $A$ is
bounded away from zero) and defect chaos (when the phase of $A$ displays
singularities where $A\!=\!0$).  The transition from phase to defect chaos
can either be hysteretic or continuous; in the former case, it is
referred to as $L_3$, in the latter as $L_1$ (Fig.~\ref{figPD}).  Despite
intensive studies
\cite{chao1,bazhenov,chate,saka,egolf,miguel,torcini,saar,mvh}, the
phenomenology of the CGLE and in particular its ``phase''-diagram
\cite{chao1,chate} are far from being understood.  Moreover, it is under
dispute whether the $L_1$ transition is sharp, and whether a pure
phase-chaotic ({\it i.e.}  defect-free) state can exist in the thermodynamic
limit \cite{egolf}.

It is the purpose of this paper to elucidate these issues by presenting the
mechanism which creates defects in transient phase chaotic states. Our
analysis consists of four parts: {\it (i)} We describe a family of Modulated
Amplitude Waves (MAWs), {\it i.e.}, pulse-like coherent structures with a
characteristic spatial period $P$.  {\it (ii)} A bifurcation analysis of these
MAWs reveals that their range of existence is limited by a saddle-node (SN)
bifurcation.  For all $c_1,c_3$ within a certain range, we define $P_{SN}$ as
the period of the MAW for which this bifurcation occurs.  {\it (iii)} We show
that for $P\!>\!P_{SN}$, {\it i.e.}, beyond the SN bifurcation, near-MAW
structures display a nonlinear evolution to defects.  It is found that, in
phase chaos, near-MAWs with various $P$'s are created and annihilated
perpetually.

\noindent {\em The transition to defect chaos takes place when near-MAWs with
$P>P_{SN}$ occur in a phase chaotic state}.  {\it (iv)} Finally, instabilities
to splitting of {\it resp.}  interaction between MAWs are identified as the
relevant processes which locally decrease {\it resp.} increase $P$ in phase
chaos.  We will argue that the SN curve for $P\rightarrow \infty$ is a lower
bound (see Fig. \ref{figPD}) for the transition from phase chaos to defect
chaos.

From a general viewpoint, our analysis shows that there is no collective
behavior that drives the transition. Instead, strictly local fluctuations
drive local structures beyond their SN bifurcation and create defects.

\begin{figure} \vspace{-0.1cm}
 \epsfxsize=1.\hsize \mbox{\hspace*{-.06 \hsize} \epsffile{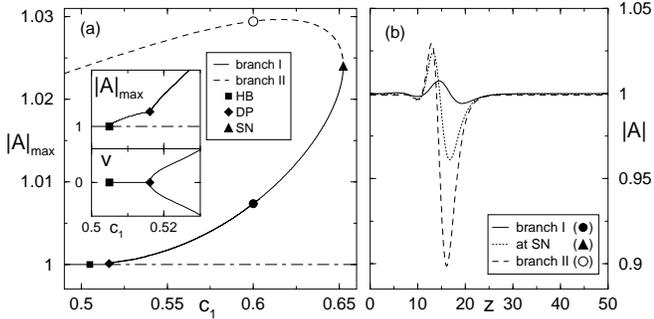} }
\caption[]{(a) Example of the bifurcation diagram of MAWs for $c_3\!=\!2.0,
P\!=\!50$ (see text). The inset illustrates the drift pitchfork bifurcation
($v\!=\!0$ branch not shown beyond bifurcation). (b) MAW profiles at lower
(full circle) and upper (open circle) branch, and at the SN
(triangle).}\label{fig2}
\end{figure}

\begin{figure} \vspace{-0.1cm}
 \epsfxsize=.98\hsize \mbox{\hspace*{-.07 \hsize}  \epsffile{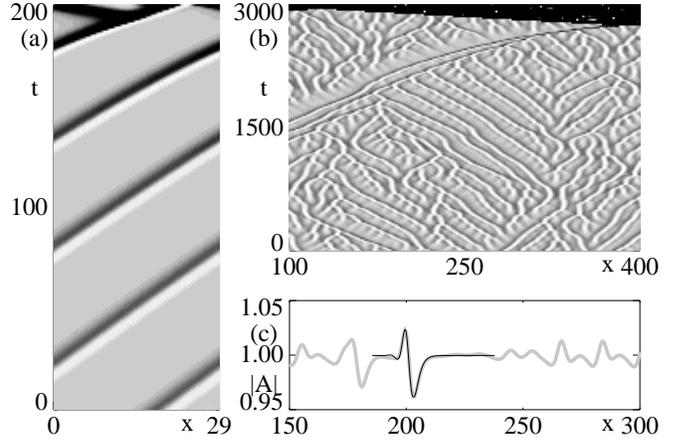} 
}
\vspace{2mm}
\caption[]{(a) Grey-scale plot of $|A|$ (black: $|A|\!\rightarrow\! 0$)
showing the nonlinear evolution of a near-MAW to defects when
$L\!=\!29\!>\!P_{SN}\!=\!26.8$ ($c_1\!=\!0.7, c_3\!=\!2$). (b) MAWs and
defect generation in a transient phase chaotic state ($c_1\!=\!0.66,
c_3\!=\!2.0$); a transient of $t\!\approx\!10^4$ is not shown. (c) Comparison
of a MAW (black) with a snapshot from a phase chaotic state (grey)
($c_1\!=\!0.66, c_3\!=\!2.0$).  }\label{figsn}
\end{figure}

{\em (i) MAWs as coherent structures -} By coherent structures we mean
uniformly propagating structures of the form \cite{torcini,saar,mvh}
\begin{equation}
A(x,t)= a(x-vt) e^{i \phi (x-vt)} e^{-i \omega t}~,
\end{equation} 
where $a$ and $\phi$ are real-valued functions of $z\!:=\!x - vt$. Such
structures play an important role in various dynamical regimes of the CGLE
\cite{miguel,torcini,saar,mvh}.  The substitution of Ansatz (2) into the CGLE
leads to a set of three coupled ODEs for $a$, $b\!=\!da/dz$ and $\psi\!=\!
d\phi/dz$ \cite{equations}.  The MAWs correspond to limit-cycles of these
ODEs, or equivalently, spatially periodic solutions of the CGLE.  The MAWs
occur in a two parameter family which we choose to parametrize by their
spatial period $P$ and their average phase gradient $\nu \!:=\! 1/P \int_0^P
dz \psi$.  Some examples of MAWs are shown in Fig.~\ref{fig2}b and
Fig.~\ref{figsn}.  Only solutions for which $\nu\!=\!0$ are considered here;
the reason for this will be discussed in {\it (iii)}. To compute the MAWs and
their bifurcations, we have used the software package AUTO94 \cite{Auto94} to
solve the ODEs for fixed $P$ and $\nu$.

{\em (ii) MAW range of existence -} MAWs with $\nu\!\neq\!0$ bifurcate from
unstable plane waves in the CGLE.  We focus on the $\nu\!=\!0$ case, {\it
i.e.}, on the homogeneous oscillation $A(x,t) \!=\! e^{i c_3 t}$.  
This solution becomes Benjamin-Feir (BF) unstable at $c_1 c_3\!=\!1$,
beyond which all plane waves are unstable (Benjamin-Feir-Newell (BFN)
criterion) \cite{CH}. 
In the ODEs, the fixed point $(a,b,\psi)\!=\!(1,0,0)$ that corresponds
to the homogeneous solution undergoes a Hopf bifurcation (HB) upon
increasing $c_1$ and $c_3$. For infinite $P$ the Hopf bifurcation
occurs for $c_1 c_3\!=\!1$, while for smaller $P$ the Hopf bifurcation
occurs for larger $c_1$ and $c_3$. The sequence of bifurcations for
{\em fixed} $P\!=\!50$ is illustrated in Fig.~\ref{fig2}a. The square
symbol denotes the Hopf bifurcation, and the resulting solutions
have drifting velocity $v\!=\!0$.  Via a secondary drift pitchfork (DP) 
bifurcation \cite{DwightDP} (diamond) the MAWs acquire $v\! \ne\!  0$.  For the
relevant parameters, {\it i.e.}, sufficiently small $\nu$ and large $P$, both
bifurcations are supercritical \cite{janiaud}; the amplitude modulations grow
away from these bifurcations.  The MAWs 

\noindent undergo a saddle-node (SN) bifurcation
(triangle) when $c_1$ or $c_3$ are sufficiently increased. The upper branch
returns far back into the BF stable region of the CGLE; the recently
discovered ``homoclinic holes'' \cite{mvh} are MAWs of this upper branch in
the limit $P \rightarrow \infty$. The spatial profiles of MAWs on the upper
(II) and lower (I) branches and SN are shown in Fig.~\ref{fig2}b.

The SN curves in the $c_1\!\!-\!\!c_3$ parameter plane have been computed for
 various spatial periods $P$. For given parameters $c_1$ and $c_3$, we define
 $P_{SN}$ as the period for which a saddle-node bifurcation occurs.  We find,
 roughly, that for larger $P$ this SN occurs for smaller values of $c_1,c_3$
 (see Fig.~\ref{figPD}).

To summarize: a family of coherent, periodic MAW solutions of the CGLE has
been obtained.  The range of existence of these solutions is limited by a SN
bifurcation for large $c_1,c_3$.

{\em (iii) Beyond the Saddle Node -} In Fig.~\ref{figsn} the relevance of the
SN for defect generation is illustrated.  In Fig.~\ref{figsn}a we show the
time evolution of a MAW-like initial condition in a periodic system of size $L
>P_{SN} $. While for $L<P_{SN}$ we obtain coherent MAWs, for $L>P_{SN}$
incoherent dynamics occurs: the amplitude modulation and drifting velocity
grow until defects are formed.  Extensive tests show that defects are always
generated for MAW-like initial conditions when $L>P_{SN}$. In
Fig.~\ref{figsn}b,c the relevance of this defect generating mechanism for
chaotic states is illustrated in a {\em large} system of size $L\!=\!512$ with
coefficients close to the $L_3$ transition.  The transient
phase chaotic state (Fig.~\ref{figsn}b) contains local structures
which can come arbitrarily close to one-period MAWs.  Fig.~\ref{figsn}c shows
a snapshot of a spatial profile of $|A|$ in a phase chaotic state; parts of
this profile can be approximated 

\begin{figure} \vspace{0cm}
 \epsfxsize=1.\hsize \mbox{\hspace*{-.05 \hsize} \epsffile{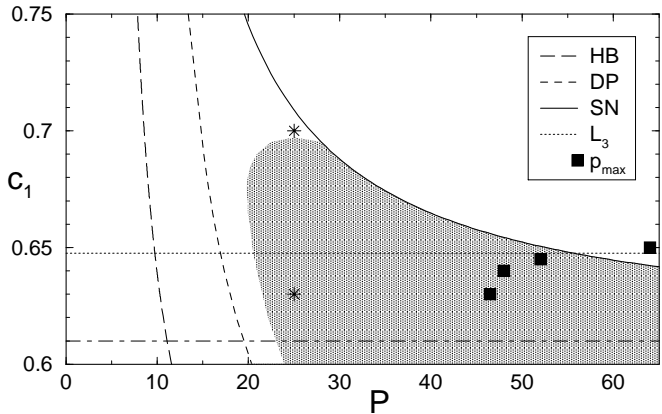} }
\caption[]{ Location of bifurcations and instabilities of MAWs as function of
$c_1$ and $P$ for $c_3 \!=\! 2.0$.  Regions unstable to splitting modes are
shaded.  For large $P$, Hopf and drift pitchfork bifurcation approach the BFN 
line and the SN curve approaches $c_1 \approx 0.61$ (dot-dashed
line). The two stars mark parameters corresponding to stability spectra shown
in Fig.~\ref{figSPEC2}. Black squares show the numerically measured maximum
peak to peak distance $p_{max}$; once these squares cross the SN curve,
defects occur. This is consistent with the numerically found location of $L_3$
(dotted line). }
\label{figSPEC1}
\end{figure}

\noindent by a MAW with appropriate $P$.  The phase
gradient $\nu$ averaged between peaks of the amplitude is always close to
zero; this is the reason why we focused on $\nu\!=\!0$ MAWs.  Defects appear
when one of these MAWs acquires a period larger than $P_{SN}$
(Fig.~\ref{figsn}b).  This illustrates the main result: the transition to
defect chaos occurs when a phase chaotic state contains pulses with peak to
peak distances larger than $P_{SN}$.

To test the generality of this picture, we have carried out extensive
numerical simulations of Eq. (1) near the transition lines $L_1$ {\it resp.}
$L_3$, adopting an integration algorithm developed in \cite{torcini}, in
systems with sizes ranging from $L\!=\!100$ to $L\!=\!5000$ and integration
times up to $5 \times 10^6$.  The distribution of peak-to-peak distances $p$
of the phase gradients has been determined.  Even though the phase chaotic
state is not everywhere MAW-like, we found that occurrences of large values of
this ``local'' $p$ were approximated well by MAW profiles.  Defects occurred
in systems with $L\ge512$ if and only if $p>P_{SN}$.  Since large $p$'s are
most ``dangerous'', the maximum value of $p$, $p_{max}$, is the relevant
quantity here.  An example of $p_{max}$ as a function of $c_1$ near $L_3$ is
shown in Fig.~\ref{figSPEC1} (squares); as soon as $p_{max}$ crosses the SN
curve, defects occur.

One may worry whether $p_{max}$ is a well-defined quantity, especially
in the thermodynamic limit. For larger system sizes and integration
times $p_{max}$ increases, however the apparent transition where
defects occur {\em shifts accordingly}.  For example, we found in our
simulations that for $c_3\!=\!2.0$, the critical value of $c_1$
approximates $0.65$, while Ref. \cite{chate} finds, for shorter
integration times, a critical value $\approx\!  0.68$.  The fact that
$p_{max}$ (slowly)

\begin{figure} \vspace{0cm}
  \epsfxsize=1.9\hsize \mbox{\hspace*{-0.08 \hsize}
    \epsffile{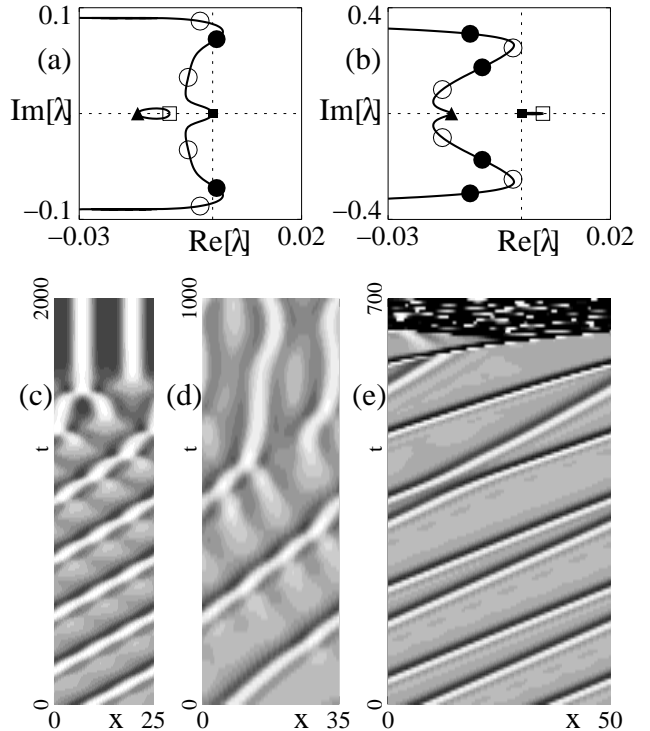} } \vspace{-1.5cm}
\caption[]{(a-b) Two typical stability spectra for $c_3\!=\!2.0$, $\nu\!=\!0$,
$P\!=\!25$ and (a) $c_1\!=\!0.63$ {\it resp.} (b) $c_1\!=\!0.70$.  Filled
symbols correspond to eigenvalues obtained for $L\!=\!P$, while open symbols
denote additional eigenvalues for $L\!=\!2P$ (symmetry modes: full square,
splitting: circles, SN: triangle, interaction: open square); the curves show
the spectrum for $L \rightarrow \infty$.  (c-d) Illustration of the splitting
instability that decreases $p$ and prevents defects to occur ($c_1 \!=\!
0.63$). For small $L$ (c) the splitting leads to a stationary pre-drift
pitchfork MAW, but for larger $L$ (d) disordered dynamics sets in.  (e) Pulse
interaction increases $p$ beyond $P_{SN}$ and leads to defects ($c_1
\!=\!0.7$).  }
\label{figSPEC2}
\end{figure}

\noindent increases for larger systems/longer times is in agreement
with earlier assertions that there is no sharp transition to defect chaos
\cite{egolf}.  We have not been able to establish an upper bound for the $p$'s
occurring in phase chaos; therefore we conjecture that the SN line for $P
\!\rightarrow\!  \infty$ provides a lower boundary for the transition from
phase to defect chaos.

{\em (iv) MAW stability - } Of course, the laminar patches that occur in MAWs
of large period are linearly unstable, and large P-MAWs have only a small
probability to occur.  To get some further insight in the behavior of MAWs, we
have calculated the linear stability properties of the MAWs.  We start with a
system of size $L\!=\!P$ and periodic boundary conditions. Both MAW branches
have neutral modes corresponding to translational and phase symmetries. The
eigenvalue associated with the SN is positive for solutions on branch II and
negative for MAWs on branch I. Apart from these 3 purely real eigenvalues, the
stability spectrum consists of pairs of complex conjugate eigenvalues.

In what follows the lower branch I is considered exclusively. For small enough
$P$, all eigenvalues $\lambda_i \!<\!0$, but when we increase $P$, MAWs become
unstable to finite wavenumber perturbations.  By using a Bloch Ansatz, we
extended the stability analysis to systems with $n$ identical pulses ($L\!
=\! nP$). For $n\!>\!1$, new instabilities may appear. The shape of these
eigenmodes suggests that the instabilities lead to splitting of {\it resp.}
interaction between adjacent MAWs; a nonlinear analysis confirms this.  These
instabilities are the relevant processes which locally decrease {\it resp.}
increase $p$, thus inhibiting or enhancing the generation of defects.  The
splitting and interaction mechanism is very similar to the cell splitting and
instabilities one encounters in the Kuramoto-Sivashinsky equation \cite{KS}.

The results of the stability analysis are summarized in Fig.~\ref{figSPEC1}
and \ref{figSPEC2}. It is important to stress here that there is no
qualitative difference between the behavior of MAWs near the $L_3$ and the
$L_1$ transition. 

The eigenvalues with largest real part on the connected
curve in Fig.~\ref{figSPEC2}a,b correspond to ``splitting'' modes;
Fig.~\ref{figSPEC2}c,d displays the nonlinear evolution that occurs when this
mode is unstable. Clearly, this instability tends to reduce the spatial
periods $p$ and prevents MAWs to cross the SN boundary. Above a critical value
for $c_1$ ($c_3$) the splitting modes are stable for all $P$
(Fig.~\ref{figSPEC1}). In this case the period of the MAWs can grow until
$P\!>\!P_{SN}$ is reached and defects are created. 

The eigenvalues labeled by open squares in Fig.~\ref{figSPEC2}a,b describe
interaction between subsequent peaks that occur for $n\!>\!1$ \cite{michal}.
These interaction modes are mainly active for small $P$ (typically $P < 20$).
They cause instability of periodic MAWs and lead to local increase of the peak
to peak distance $p$; Fig.~\ref{figSPEC2}e shows the nonlinear evolution in
such a case.

{\em Conclusion -} We have presented a systematic study of modulated
amplitude waves (MAWs) in the complex Ginzburg-Landau equation (CGLE).
These periodic coherent structures originate in supercritical
bifurcations due to the BF instability of the CGLE. MAW existence is
bounded by saddle-node bifurcations towards large $c_1, c_3$.
Approaching the transition from phase to defect chaos, near-MAWs with
large $P$ occur in phase chaos.  Defects are generated if the period
of these MAWs becomes larger than $P_{SN}$.  This scenario is valid
for both the $L_1$ and $L_3$ transition.  Indications have been given
in favor of the existence of the phase turbulent regime even in the
thermodynamic limit. Altogether, our study leaves little space for
doubt that the transition from phase chaos to defect chaos in the CGLE
is governed by coherent structures and their bifurcations.

It is a pleasure to acknowledge discussions with H. Chat\'e and L. Kramer.  AT
and MB are grateful to ISI Torino for providing a pleasant working environment
during the Workshop on ``Complexity and Chaos'' in October 1999.  MGZ is
supported from a post-doctoral grant of the MEC (Spain). MvH acknowledges
financial support from the EU under contract ERBFMBICT 972554.

\end{multicols}


\begin{references}
\bibitem{equations} The three ODE system reads: $ a_z \!=\! b ; \\ b_z \!=\! \psi^2 a
- \gamma^{-1} [(1+c_1 \omega)a+v(b+c_1\psi a)+(c_1 c_3-1)a^3]; \\ \psi_z \!=\!
-2b\psi/a+\gamma^{-1} [c_1-\omega+v(c_1b/a-\psi)-(c_1+c_3) a^2].$
Abbreviations used are $\gamma \!=\! 1 + c_1^2$, $a_z \!=\! da/dz, b_z \!=\! db/dz$ and
$\psi_z \!=\! d\psi/dz$.
\end{references}
\end{document}